\documentclass[aps,prl,twocolumn,superscriptaddress,showpacs,preprintnumbers,amsmath,amssymb]{revtex4}

\usepackage{graphicx} % Include figure files
\usepackage{dcolumn}  % Align table columns on decimal point
\usepackage{amsmath}

\graphicspath{{ps}}

\begin{document}

%\preprint{ \vbox{        \hbox{\hfil \today , version 4.2}
%                        \hbox{\hfil Intended for {\it PRL}}
                         %\hbox{\hfil Author: S.~U.~Kataoka, K.~Miyabayashi}
                         %\hbox{\hfil Committee: T.~Higuchi~(Chair), F.~Ronga, E.~Barberio}
%}}

\title{ \quad\\[0.5cm]  Study of Time-dependent $CP$ Violation 
in $B^0 \rightarrow J/\psi \pi^0$ Decays}

%\input author.tex
%%%
% Author list
%%%
\affiliation{Budker Institute of Nuclear Physics, Novosibirsk}
\affiliation{Chiba University, Chiba}
\affiliation{Chonnam National University, Kwangju}
%%%\affiliation{Chuo University, Tokyo}
\affiliation{University of Cincinnati, Cincinnati, Ohio 45221}
\affiliation{University of Frankfurt, Frankfurt}
%%%\affiliation{Gyeongsang National University, Chinju}
\affiliation{University of Hawaii, Honolulu, Hawaii 96822}
\affiliation{High Energy Accelerator Research Organization (KEK), Tsukuba}
\affiliation{Hiroshima Institute of Technology, Hiroshima}
\affiliation{Institute of High Energy Physics, Chinese Academy of Sciences, Beijing}
\affiliation{Institute of High Energy Physics, Vienna}
\affiliation{Institute for Theoretical and Experimental Physics, Moscow}
\affiliation{J. Stefan Institute, Ljubljana}
\affiliation{Kanagawa University, Yokohama}
\affiliation{Korea University, Seoul}
%%%\affiliation{Kyoto University, Kyoto}
\affiliation{Kyungpook National University, Taegu}
\affiliation{Swiss Federal Institute of Technology of Lausanne, EPFL, Lausanne}
\affiliation{University of Ljubljana, Ljubljana}
\affiliation{University of Maribor, Maribor}
\affiliation{University of Melbourne, Victoria}
\affiliation{Nagoya University, Nagoya}
\affiliation{Nara Women's University, Nara}
\affiliation{National Central University, Chung-li}
%%%\affiliation{National Kaohsiung Normal University, Kaohsiung}
\affiliation{National United University, Miao Li}
\affiliation{Department of Physics, National Taiwan University, Taipei}
\affiliation{H. Niewodniczanski Institute of Nuclear Physics, Krakow}
\affiliation{Nihon Dental College, Niigata}
\affiliation{Niigata University, Niigata}
\affiliation{Osaka City University, Osaka}
\affiliation{Osaka University, Osaka}
\affiliation{Panjab University, Chandigarh}
\affiliation{Peking University, Beijing}
\affiliation{Princeton University, Princeton, New Jersey 08545}
%%%\affiliation{RIKEN BNL Research Center, Upton, New York 11973}
%%%\affiliation{Saga University, Saga}
\affiliation{University of Science and Technology of China, Hefei}
\affiliation{Seoul National University, Seoul}
\affiliation{Sungkyunkwan University, Suwon}
\affiliation{University of Sydney, Sydney NSW}
\affiliation{Tata Institute of Fundamental Research, Bombay}
\affiliation{Toho University, Funabashi}
\affiliation{Tohoku Gakuin University, Tagajo}
\affiliation{Tohoku University, Sendai}
\affiliation{Department of Physics, University of Tokyo, Tokyo}
\affiliation{Tokyo Institute of Technology, Tokyo}
\affiliation{Tokyo Metropolitan University, Tokyo}
\affiliation{Tokyo University of Agriculture and Technology, Tokyo}
%%%\affiliation{Toyama National College of Maritime Technology, Toyama}
\affiliation{University of Tsukuba, Tsukuba}
%%%\affiliation{Utkal University, Bhubaneswer}
\affiliation{Virginia Polytechnic Institute and State University, Blacksburg, Virginia 24061}
\affiliation{Yonsei University, Seoul}
  \author{S.~U.~Kataoka}\affiliation{Nara Women's University, Nara} % Nara
  \author{K.~Abe}\affiliation{High Energy Accelerator Research Organization (KEK), Tsukuba} % KEK
  \author{K.~Abe}\affiliation{Tohoku Gakuin University, Tagajo} % TohokuGakuin
% \author{N.~Abe}\affiliation{Tokyo Institute of Technology, Tokyo} % TIT
  \author{I.~Adachi}\affiliation{High Energy Accelerator Research Organization (KEK), Tsukuba} % KEK
  \author{H.~Aihara}\affiliation{Department of Physics, University of Tokyo, Tokyo} % Tokyo
  \author{M.~Akatsu}\affiliation{Nagoya University, Nagoya} % Nagoya
  \author{Y.~Asano}\affiliation{University of Tsukuba, Tsukuba} % Tsukuba
% \author{T.~Aso}\affiliation{Toyama National College of Maritime Technology, Toyama} % Toyama
% \author{V.~Aulchenko}\affiliation{Budker Institute of Nuclear Physics, Novosibirsk} % BINP
  \author{T.~Aushev}\affiliation{Institute for Theoretical and Experimental Physics, Moscow} % ITEP
% \author{T.~Aziz}\affiliation{Tata Institute of Fundamental Research, Bombay} % Tata
  \author{S.~Bahinipati}\affiliation{University of Cincinnati, Cincinnati, Ohio 45221} % Cincinnati
  \author{A.~M.~Bakich}\affiliation{University of Sydney, Sydney NSW} % Sydney
% \author{Y.~Ban}\affiliation{Peking University, Beijing} % Peking
% \author{M.~Barbero}\affiliation{University of Hawaii, Honolulu, Hawaii 96822} % Hawaii
% ...
% Ljubljana
% \author{D.~Z\"urcher}\affiliation{Swiss Federal Institute of Technology of Lausanne, EPFL, Lausanne} % Lausanne
\collaboration{The Belle Collaboration}
%%%%%%%%%%%%%%%

\begin{abstract}
We report a measurement of $CP$ asymmetry parameters in the 
decay $B^0(\overline B{}^0) \rightarrow J/\psi \pi^0$,
which is governed by the $b \rightarrow c\bar{c}d$ transition.
The analysis is based on 
a 140 fb$^{-1}$ data sample accumulated at
the $\Upsilon(4S)$ resonance
by the Belle detector 
at the KEKB asymmetric-energy $e^+ e^-$ collider.
We fully reconstruct one neutral $B$ meson in the
$J/\psi \pi^0$ final state.   The accompanying
$B$ meson flavor is identified by its decay products.
From the distribution of proper time intervals between the two
$B$ decays, we obtain the following $CP$-violating parameters: 
$\ S_{J/\psi \pi^0}$ = $-0.72 \pm0.42 (\mbox{stat}) \pm 0.09(\mbox{syst})$ and 
$\ A_{J/\psi \pi^0}$ = $-0.01 \pm0.29 (\mbox{stat}) \pm 0.03(\mbox{syst})$.
\end{abstract}


\maketitle

%%%% keep the final version single-spaced
\tighten

{\renewcommand{\thefootnote}{\fnsymbol{footnote}}}
\setcounter{footnote}{0}

%%%%%%%
%\section{Introduction}
%%%%%%
In the Standard Model (SM), the Kobayashi-Maskawa (KM)
quark-mixing matrix~\cite{KM}
has an irreducible complex phase that gives rise to $CP$ violation in weak
interactions.
In particular, the SM predicts large $CP$-violating asymmetries in the
time-dependent rates of $B^0$ and $\overline B{}^0$ decays into a common
$CP$ eigenstate $f_{CP}$~\cite{sanda}.
In the decay chain $\Upsilon(4S)\rightarrow B^0 \overline B{}^0
\rightarrow f_{CP}f_{\rm tag}$,
where one of the $B$ mesons decays at time $t_{CP}$ to a final state
$f_{CP}$
and the other decays at time $t_{\rm tag}$ to a final state
$f_{\rm tag}$ that distinguishes between $B^0$ and $\overline B{}^0$,
the decay rate has a time dependenct probability
given by~\cite{CPVrev}
\begin{eqnarray}
\label{eq:psig}
{\cal P}(\Delta t) =
\frac{e^{-|\Delta t|/{\tau_{B^0}}}}{4{\tau_{B^0}}}
\bigg\{1 + q
\Big[ {\cal S}_{f_{CP}} \sin(\Delta m_d \Delta t) \nonumber \\
   + {\cal A}_{f_{CP}} \cos(\Delta m_d \Delta t)
\Big]
\bigg\},
\end{eqnarray}
where $\tau_{B^0}$ is the $B^0$ lifetime, $\Delta m_d$ is
the mass difference between the two $B^0$ mass
eigenstates, $\Delta{t}$ $\equiv$ $t_{CP}$ $-$ $t_{\rm tag}$, and
the $b$-flavor $q$ = +1 ($-1$) when the tagging $B$ meson
is a $B^0$ ($\overline B{}^0$).
The $CP$-violating parameters ${\cal S}_{f_{CP}}$ and
${\cal A}_{f_{CP}}$  are given by
\begin{equation}
{\cal S}_{f_{CP}} \equiv \frac{2\Im(\lambda)}{|\lambda|^2+1}, \qquad
{\cal A}_{f_{CP}} \equiv \frac{|\lambda|^2-1}{|\lambda|^2+1},
\end{equation}
where $\lambda$ is a complex
parameter that depends on both the $B^0 \overline B{}^0$
mixing and on the amplitudes for $B^0$ and $\overline B{}^0$ decay to $f_{CP}$.
To a good approximation in the SM,
$|\lambda|$ is equal to the absolute value
of the ratio of the $\overline B{}^0 \rightarrow f_{CP}$ to
$B^0 \rightarrow f_{CP}$ decay amplitudes.

$CP$ violation in neutral $B$ meson decays involving the
$b \to c \bar{c} s$ transition has been established through  measurements of
the $CP$-violating  parameter $\sin2\phi_1$ by
the Belle~\cite{phi1_Belle} and BaBar~\cite{phi1_BaBar} collaborations.
The SM predicts ${\cal S}_{f_{CP}} = -\xi_f \sin 2\phi_1$,
where $\xi_f = +1 (-1)$
corresponds to  $CP$-even (-odd) final states; and ${\cal A}_{f_{CP}} =0$
(or equivalently $|\lambda| = 1$) 
for tree diagrams in both $b \rightarrow c\bar{c}s$ and
$b \rightarrow c\bar{c}d$.
In contrast with the
$b \rightarrow c\bar{c}s$ case, however,  
tree and penguin amplitudes all contribute to $b \rightarrow c\bar{c}d$
to the same order in the sine of the Cabibbo angle.
Therefore, if penguin or other contributions are
substantial, a precision measurement of the
time-dependent $CP$ asymmetry in $b \rightarrow c \bar{c} d$ may reveal
values for ${\cal S}_{J/\psi\pi^0}$ and ${\cal A}_{J/\psi\pi^0}$ 
that differ from the values for $b \rightarrow c\bar{c}s$.
Measurements of $CP$ asymmetries in
$b \rightarrow c\bar{c}d$ transition-induced $B$ decays such as
$B^0 \rightarrow J/\psi\pi^0$ thus play an important role 
in probing these one-loop diagrams.

A study of $CP$ asymmetry in $B^0 \rightarrow J/\psi\pi^0$ decays
has been reported by the BaBar collaboration~\cite{psipi0CP_BaBar}
based on 81 fb$^{-1}$.
In this paper
we report a measurement of time-dependent $CP$-violating parameters
in $B^0 \rightarrow J/\psi\pi^0$ decays using 
a data sample of 140 fb$^{-1}$
(corresponding to 15.2$\times 10^7$ $B\overline B{}$ pairs) collected at
the $\Upsilon(4S)$ resonance with the Belle detector~\cite{Belle}
at the KEKB asymmetric-energy $e^+e^-$ (3.5~GeV on 8.0~GeV) collider~\cite{KEKB}.
The $\Upsilon(4S)$ is produced
with a Lorentz boost of $\beta\gamma=0.425$ nearly along
the $z$-axis defined as anti-parallel to the positron beam.  
Since the $B^0$ and $\overline{B}{}^0$ mesons are nearly at 
rest in the $\Upsilon(4S)$ center-of-mass system (cms),
$\Delta t$ can be determined from $\Delta z$,
the displacement in $z$
between the $f_{CP}$ and $f_{\rm tag}$ decay vertices:
$\Delta t \simeq \Delta z/\beta\gamma c
\equiv (z_{CP} - z_{\rm tag})/\beta\gamma c$

The Belle detector is a large-solid-angle magnetic
spectrometer that
consists of a three-layer silicon vertex detector (SVD),
a 50-layer central drift chamber (CDC), an array of
aerogel threshold Cherenkov counters (ACC),
a barrel-like arrangement of time-of-flight
scintillation counters (TOF), and an electromagnetic calorimeter
comprised of 
CsI(Tl) crystals (ECL) located inside
a superconducting solenoid coil that provides a 
1.5 T
magnetic field.  An iron flux-return located outside of
the coil is instrumented to detect $K_L^0$ mesons and to identify
muons (KLM).  
The selection of hadronic events is described elsewhere~\cite{Schrenk}.

%-----------
%\subsection{Reconstruction of $J/\psi$ mesons}
%----------
$J/\psi$ mesons are reconstructed via their decays into 
oppositely-charged lepton pairs ($e^+e^-$ or $\mu^+\mu^-$).   
Both lepton tracks must be positively identified~\cite{Schrenk}.
In the $e^+e^-$ mode, ECL clusters
that have no associated charged tracks and 
are within 50~mrad of the track's initial momentum vector are
included in the calculation of the invariant mass ($M_{ee(\gamma)}$),
in order to include photons radiated
from electrons/positrons.
The invariant masses of $e^+e^-(\gamma)$ and $\mu^+\mu^-$ combinations 
are required to fall within the ranges  
$-150\mbox{~MeV}/c^2 < (M_{ee(\gamma)} - M_{J/\psi}) < 36\mbox{~MeV}/c^2$ and
$-60\mbox{~MeV}/c^2 < (M_{\mu\mu} - M_{J/\psi}) < 36\mbox{~MeV}/c^2$, respectively.
Here $M_{J/\psi}$ denotes the world average of the $J/\psi$ mass~\cite{PDG}.

%----------
%\subsection{Reconstruction of $\pi^0$ mesons}
%----------
Photon candidates are selected from clusters of up to 5$\times$5 crystals
in the ECL.
Each photon candidate is required to have no associated charged track,
and a cluster shape that is consistent with an electromagnetic shower.
To select $\pi^0 \rightarrow \gamma \gamma$ decay candidates
for the $B^0 \rightarrow J/\psi\pi^0$ mode,
the energy of each photon is required to exceed 50~MeV
in the ECL barrel ($32^{\circ} < \theta < 129^{\circ}$) or
100~MeV in the forward ($17^{\circ} < \theta < 32^{\circ}$)
or backward ($129^{\circ} < \theta < 150^{\circ}$) endcaps,
where $\theta$ denotes the polar angle with respect to the $z$-axis.
Neutral pion candidates are formed
from photon pairs that have an invariant mass
in the range 0.118~GeV/$c^2$ to 0.150~GeV/$c^2$.

$J/\psi$ and $\pi^0$ candidates are combined to form $B$ candidates.
The $B$ candidate selection is carried out using two observables
in the $\Upsilon(4S)$ cms:
the beam-energy constrained mass
$M_{\rm bc} \equiv \sqrt{ E_{\rm beam}^2 - (\sum \vec{p_i})^2} $
and the energy difference $\Delta E \equiv \sum E_i - E_{\rm beam}$,
where $E_{\rm beam}=\sqrt{s}/2$ is the cms beam energy,
and $\vec{p_i}$ and $E_i$ are the cms three-momenta and energies of
the $B$ meson decay products, respectively.
In this calculation, $\vec{p_i}$ and $E_i$ are obtained after refitting with  
vertex and mass constraints for the $J/\psi$ di-lepton decays
and a mass constraint for the $\pi^0 \rightarrow \gamma \gamma$ decays
in order to improve the $\Delta E$ and $M_{\rm bc}$ resolutions.
The $B$ meson signal region is defined as
$5.27\mbox{~GeV}/c^2 < M_{\rm bc} < 5.29\mbox{~GeV}/c^2$ and
$-0.10\mbox{~GeV} < \Delta E < 0.05\mbox{~GeV}$.
The lower bound of $\Delta E$ 
is chosen to
accommodate the negative $\Delta E$ tail 
due to shower leakage associated with
$\pi^0$,
and avoid the background at $\Delta E \sim -0.2$~GeV
due to $B^0 \rightarrow J/\psi ~K_S(K_S \rightarrow \pi^0 \pi^0)$ events. 
The number of reconstructed $B^0 \rightarrow J/\psi \pi^0$ candidates is
103.

%%%%
%\section{Flavor tagging and vertexing}
%%%%
Charged leptons, kaons, pions, and $\Lambda$ baryons
that are not associated with the reconstructed
$B^0 \rightarrow J/\psi \pi^0$  candidate
are used to identify the $b$-flavor of the accompanying $B$ meson,
which decays into $f_{\rm tag}$.
Based on the measured properties of these tracks, two parameters,
$q$ and $r$, are assigned to each event.
The first, $q$, has the discrete value $+1$~($-1$)
when the tag-side $B$ meson is more likely to be a $B^0$~($\overline B{}^0$).
The parameter $r$ is an event-by-event MC-determined
flavor-tagging quality factor that ranges
from $r=0$ for no flavor discrimination
to $r=1$ for an unambiguous flavor assignment.
It is used only to sort data into six intervals of $r$,
according to the estimated flavor purity.
The wrong-tag probabilities, $w_l$ and the difference
in wrong-tag probabilities between $B^0$ and $\overline B{}^0$,
$\Delta w_l$ ($l=1,6$) are 
fixed using a data sample of self-tagged $B^0$ decay modes.
The wrong tag fractions 
for each $r$ interval that are used in the final fit
are given elsewhere~\cite{Belle_sin2phi1_2003}.

\begin{figure}[htb]
\includegraphics[width=0.5\textwidth]{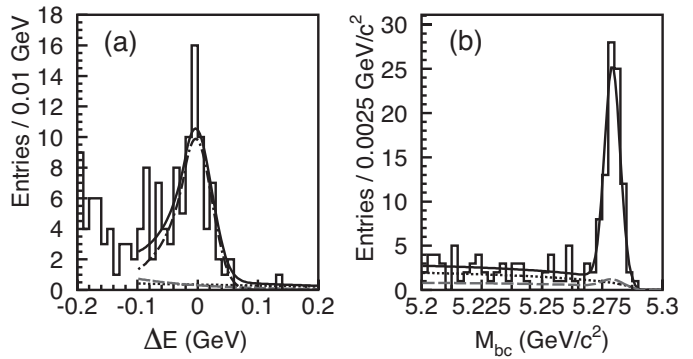}
\caption{(a) $\Delta E$ distribution
for events in the $M_{\rm bc}$ signal region
and (b) $M_{\rm bc}$ distribution
for events in the $\Delta E$ signal region,
for $B^0 \rightarrow J/\psi \pi^0$ candidates.
The superimposed curves show fitted contributions from
signal (dot-dashed line, plotted only in $\Delta E$
distribution), $B \rightarrow J/\psi X$ background (dotted line), 
combinatorial background (dashed line) and the sum of all the 
contributions (solid line).}  
%See text for further details.}
\label{mbc-deltae}
\end{figure}

%%%%
% vertexing.
%%%%%
The decay vertices of $B$ mesons are reconstructed using
tracks that have 
associated SVD hits.
Each vertex position is required to be consistent with
the interaction point profile smeared by the average transverse $B$ meson 
decay length~\cite{resol_nim}.
The vertex position for the $B^0 \rightarrow J/\psi \pi^0$ decay
is reconstructed using lepton tracks from the $J/\psi$.
The $f_{\rm tag}$ vertex is determined from all  well-reconstructed tracks 
except those used for $B^0 \rightarrow J/\psi \pi^0$ reconstruction and 
tracks that form a  $K^0_S$ candidate. 
After flavor tagging and vertex reconstruction, 
91 of the 103 $B^0 \rightarrow J/\psi \pi^0$ candidates remain.
The $\Delta E$ and $M_{\rm bc}$ distributions for the candidate events 
are shown in Fig. \ref{mbc-deltae}.

%--------- DeltaE and Mbc plots  ------
%----------------------------------------------------------

%%%%%
%\section{Signal probability}
%%%%%
To assign an event-by-event signal probability for use in
the maximum-likelihood fit to extract the $CP$-violating parameters,
we determine event distribution functions in the $\Delta E$--$M_{\rm bc}$ plane
for both signal and background.
The signal distribution is
modeled with a two-dimensional function which is 
Gaussian in $M_{\rm bc}$ and a Crystal Ball line shape~\cite{CBlineshape}
in $\Delta E$.  
Since the number of candidate events is not sufficient to obtain
precisely the shape parameters of these functions,
those are determined
from MC simulation and held fixed in the fit, 
while the overall signal yield is allowed to float.
The background  distribution is studied 
using a large sample of MC events along with events outside 
the signal region.
We split the backgrounds into two categories: $B$ decays containing 
a $J/\psi$ ($B \rightarrow J/\psi X$) and combinatorial
background to which random combinations of particles in
$B\overline B{}$ decays and continuum events contribute.
According to a MC study, the $B \rightarrow J/\psi X$ background forms
a small peak in the $M_{\rm bc}$ projection.
Therefore we parametrize this contribution with a sum of a 
Gaussian and
a phase-space-like background function (ARGUS function)~\cite{ARGUSBG}
in $M_{\rm bc}$ %SK direction 
and an exponential function for $\Delta E$.
The amount of $B \rightarrow J/\psi X$ background is estimated by
the MC study and fixed in the fit~\cite{inclusivepsi}.
For the combinatorial background, we use a linear function
for $\Delta E$ and an ARGUS function for $M_{\rm bc}$.
From the fit, the purity of the $B^0 \rightarrow J/\psi \pi^0$ candidates
is estimated to be (86$\pm$11)\%.

%%%%%
%\section{The unbinned maximum likelihood fit}
%%%%
We determine ${\cal S}_{J/\psi \pi^0}$ and ${\cal A}_{J/\psi \pi^0}$
by performing
an unbinned maximum-likelihood fit to the observed $\Delta t$ distribution.
The probability density function (PDF) expected for the signal
distribution is given by 
\begin{eqnarray}
\lefteqn{{\cal P}_{\rm sig}(\Delta t)} \nonumber \\
&=& 
 \frac{e^{-|\Delta t|/{\tau_{B^0}}}}{4{\tau_{B^0}}}
\bigg\{1 -q\Delta w_l + q(1-2w_l)\nonumber \\
& &\times \Big[ {\cal S}_{J/\psi \pi^0} \sin(\Delta m_d \Delta t)
   + {\cal A}_{J/\psi \pi^0} \cos(\Delta m_d \Delta t)
\Big]
\bigg\},\nonumber \\
& &
\end{eqnarray}
to account for the effect of incorrect flavor
assignment.
The distribution is
convolved with the
proper-time interval resolution function $R_{\rm sig}(\Delta t)$,
which takes into account the finite vertex resolution.
$R_{\rm sig}(\Delta t)$ is formed by convolving four components: 
the detector resolutions for $z_{CP}$ and
$z_{\rm tag}$, the shift in the $z_{\rm tag}$ vertex position
due to secondary tracks originating from charmed particle decays, and
the kinematic approximation that the $B$ mesons are at rest in the
cms~\cite{resol_nim}.
A small component of broad outliers in the $\Delta z$ distribution, caused
by
misreconstruction, is represented by a Gaussian function
$P_{\rm ol}(\Delta t)$.
We use the same resolution function parameters as used in 
the $\sin2\phi_1$ measurement~\cite{Belle_sin2phi1_2003}. 
We determine the following likelihood value for each
event:
\begin{eqnarray}
\lefteqn{P_i(\Delta t_i;{\cal S}_{J/\psi\pi^0},{\cal A}_{J/\psi\pi^0})}
\nonumber \\
&=& (1-f_{\rm ol})\int_{-\infty}^{\infty}\biggl[
f_{\rm sig}{\cal P}_{\rm sig}(\Delta t')R_{\rm sig}
(\Delta t_i-\Delta t')  \nonumber \\
& & \quad + f_{\rm bkg}^{J/\psi X}{\cal P}_{\rm bkg}^{J/\psi X}(\Delta t')
R_{\rm bkg}^{J/\psi X}(\Delta t_i-\Delta t') \nonumber \\
& &\quad +\; (1-f_{\rm sig}-f_{\rm bkg}^{J/\psi X})
{\cal P}_{\rm bkg}^{\rm comb}(\Delta t') \nonumber \\
& &\times R_{\rm bkg}^{\rm comb}(\Delta t_i-\Delta t')\biggr]
d(\Delta t') + f_{\rm ol} P_{\rm ol}(\Delta t_i),
\end{eqnarray}
where $f_{\rm ol}$ is the outlier fraction,
$f_{\rm sig}$ and $f^{J/\psi X}_{\rm bkg}$ are 
the signal and $B \rightarrow J/\psi X$ background probabilities 
calculated as a function
of $\Delta E$ and $M_{\rm bc}$,
${\cal P}_{\rm bkg}^{J/\psi X}(\Delta t)$ and
${\cal P}_{\rm bkg}^{\rm comb}(\Delta t)$
are  the PDFs for $B \rightarrow J/\psi X$ 
and combinatorial background events,
respectively.
${\cal P}_{\rm bkg}^{\rm comb}(\Delta t)$ is modeled as a sum of exponential and prompt components, and
 is convolved with the corresponding resolution function 
$R_{\rm bkg}^{\rm comb}$, which is modeled by a sum of two Gaussians.
The parameters in
${\cal P}_{\rm bkg}^{\rm comb} (\Delta t)$
and $R_{\rm bkg}^{\rm comb}$ are determined by a fit to
the $\Delta t$ distribution
of a background-enhanced control sample,
i.e. events not in the $\Delta E$--$M_{\rm bc}$ signal region.
${\cal P}_{\rm bkg}^{J/\psi X}(\Delta t)$ is a lifetime distribution determined  by MC simulation.
The contributions of $CP=-1$ and $+1$ components
in the $B \rightarrow J/\psi X$ background in the signal region are 
found to be almost the same
and the $CP$ violation effect is largely cancelled.
We model $R_{\rm bkg}^{J/\psi X}$ using the signal resolution function
because vertex reconstruction for both signal and $B \rightarrow J/\psi X$ background events is based on lepton tracks from $J/\psi$.
We fix  $\tau_{B^0}$ and $\Delta m_d$ at
their world-average values~\cite{PDG2003}.
The only free parameters in the final fit
are ${\cal S}_{J/\psi \pi^0}$ and ${\cal A}_{J/\psi \pi^0}$,
which are determined by maximizing the
likelihood function
\begin{equation}
{L} = \prod_iP_i(\Delta t_i;{\cal S}_{J/\psi\pi^0},{\cal
A}_{J/\psi\pi^0}),
\end{equation}
where the product runs over all events.

\begin{figure}[htb]
\includegraphics[width=0.4\textwidth]{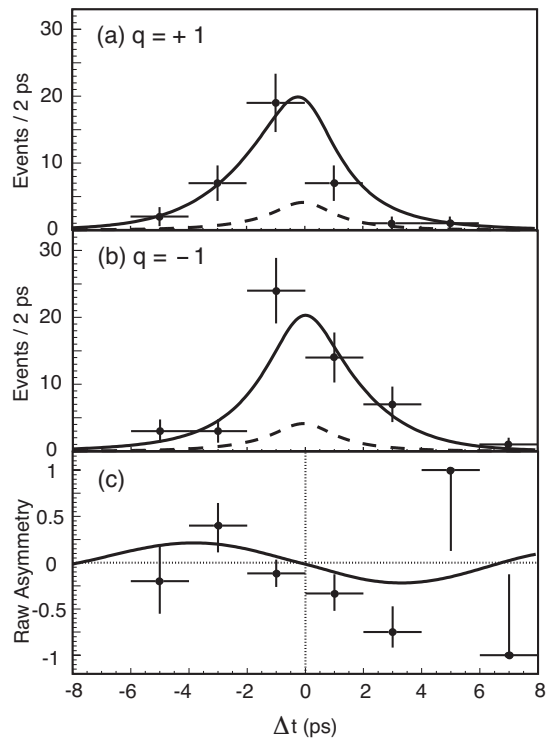}
\caption{The $\Delta t$ distributions for 
(a) $\overline B{}^0 \rightarrow J/\psi\pi^0$ ($q=+1$) and
(b) $B^0 \rightarrow J/\psi\pi^0$ ($q=-1$) candidates.
(c) The asymmetry, $A$, in each $\Delta t $ bin. 
The curve shows the result of the unbinned maximum-likelihood fit, 
and dashed curves show the background distributions.}
\label{deltat}
\end{figure}

%%%%%%%%
% Results
%%%%%%%%
A fit to the candidate events results in the $CP$-violating parameters,
\begin{eqnarray}\nonumber
{\cal S}_{J/\psi \pi^0}
&=& -0.72 \pm 0.42(\mbox{stat}) \pm 0.09(\mbox{syst}) \\
{\cal A}_{J/\psi \pi^0}
&=& -0.01 \pm 0.29(\mbox{stat}) \pm 0.03(\mbox{syst}),  
\end{eqnarray}
where the sources of systematic errors are 
described below.
Figure \ref{deltat} (a) and (b) show the $\Delta t$ distributions for  
$\overline B{}^0 \rightarrow J/\psi\pi^0$ ($q=+1$) and
$B^0 \rightarrow J/\psi\pi^0$ ($q=-1$) event samples
respectively.
Figure \ref{deltat} (c) shows the raw asymmetry in each $\Delta t $ bin
without background subtraction, which is defined by 
\begin{eqnarray}
A \equiv { { N_{q=+1} - N_{q=-1}  }\over{ N_{q=+1} + N_{q=-1} } },
\end{eqnarray}
where $N_{q=+1}$($N_{q=-1}$) is the number of observed candidates
with $q=+1$$(-1)$. The curve shows the result of 
the unbinned maximum-likelihood fit to 
the $\Delta t$ distributions.

%--------- Delta t plot goes here ------
%----------------------------------------------------------
The major systematic errors come from uncertainties
in the vertex reconstruction
($\pm 0.06$ for ${\cal S}_{J/\psi \pi^0}$ and
$\pm 0.02$ for ${\cal A}_{J/\psi \pi^0}$), 
the wrong tag fraction  
($\pm 0.03$ and $\pm 0.01$), 
the potential $CP$-violating effect in $B \rightarrow J/\psi X$ background
($\pm 0.03$ and $\pm 0.01$), 
the signal probability 
($\pm 0.02$ and $\pm 0.02$), 
and possible fit bias 
($\pm 0.03$ and $\pm 0.01$).
The contributions from other systematic error sources, 
uncertainties in the background $\Delta t$ distribution, resolution function,
and physics parameters ($\tau_{B^0}$, $\Delta m_d$),
are found to be much smaller.
The quadratic sum of all the contributions %mentioned above
amounts to 
$\pm 0.09$ for ${\cal S}_{J/\psi \pi^0}$ and
$\pm 0.03$ for ${\cal A}_{J/\psi \pi^0}$.

%%%%
%DRM \section{Summary}
%%%%
In summary, we have performed a measurement of the $CP$-violating parameters
in the $B^0 \rightarrow J/\psi \pi^0$ decay.
The resultant values are 
${\cal S}_{J/\psi \pi^0}
= -0.72 \pm 0.42(\mbox{stat}) \pm 0.09(\mbox{syst})$
and 
${\cal A}_{J/\psi \pi^0}
= -0.01 \pm 0.29(\mbox{stat}) \pm 0.03(\mbox{syst})$.
These values are consistent with those
obtained for $B^0 \rightarrow J/\psi ~K_S$ and other decays governed
by $b \rightarrow c\bar{c}s$ transition, 
as expected from the SM if the tree diagram dominates.

% 
%***** Acknowledgments *****
We thank the KEKB group for the excellent
operation of the accelerator, the KEK Cryogenics   
group for the efficient operation of the solenoid,   
and the KEK computer group and the NII for valuable computing and   
Super-SINET network support.  We acknowledge support from  
MEXT and JSPS (Japan); ARC and DEST (Australia); NSFC (contract
No.~10175071, China); DST (India); the BK21 program of MOEHRD and the   
CHEP SRC program of KOSEF (Korea); KBN (contract No.~2P03B 01324,   
Poland); MIST (Russia); MESS (Slovenia); NSC and MOE (Taiwan); and DOE  (USA).


\begin{thebibliography}{99}
\bibitem{PDG2003}
In the likelihood function, we use the following $B^0\overline B{}^0$-mixing
parameter and $B$ meson lifetimes~\cite{PDG}: $\Delta m_d$ = 0.502$\pm$0.007 ps$^{-1}$,
$\tau_{B^0}$ = 1.537$\pm$0.015 ps and
$\tau_{B^+}$ = 1.671$\pm$0.018 ps.
\end{thebibliography}
\end{document}